\documentclass[twocolumn,amssymb,preprint,aps,prl,epsf]{elsart}
\newcommand{\etalnoem}{et al.}

\def\deg{^\circ}

\newcommand{\mysect}[1]{}

\def\Journal#1#2#3#4{{#1} {\bf #2} (#4) #3}

\def\PRL{\rm Phys. Rev. Lett.}
\def\PRD{{\rm Phys. Rev.} D}
\def\NIM#1#2#3{\rm Nucl. Instr. and Meth A{#1}~(#2)~#3}

\newcommand{\FigOverview}{
\begin{center}
\begin{figure}[htbp]
\centering{\includegraphics[angle=270,viewport=0 0 600 800,clip,width=1.0\linewidth]{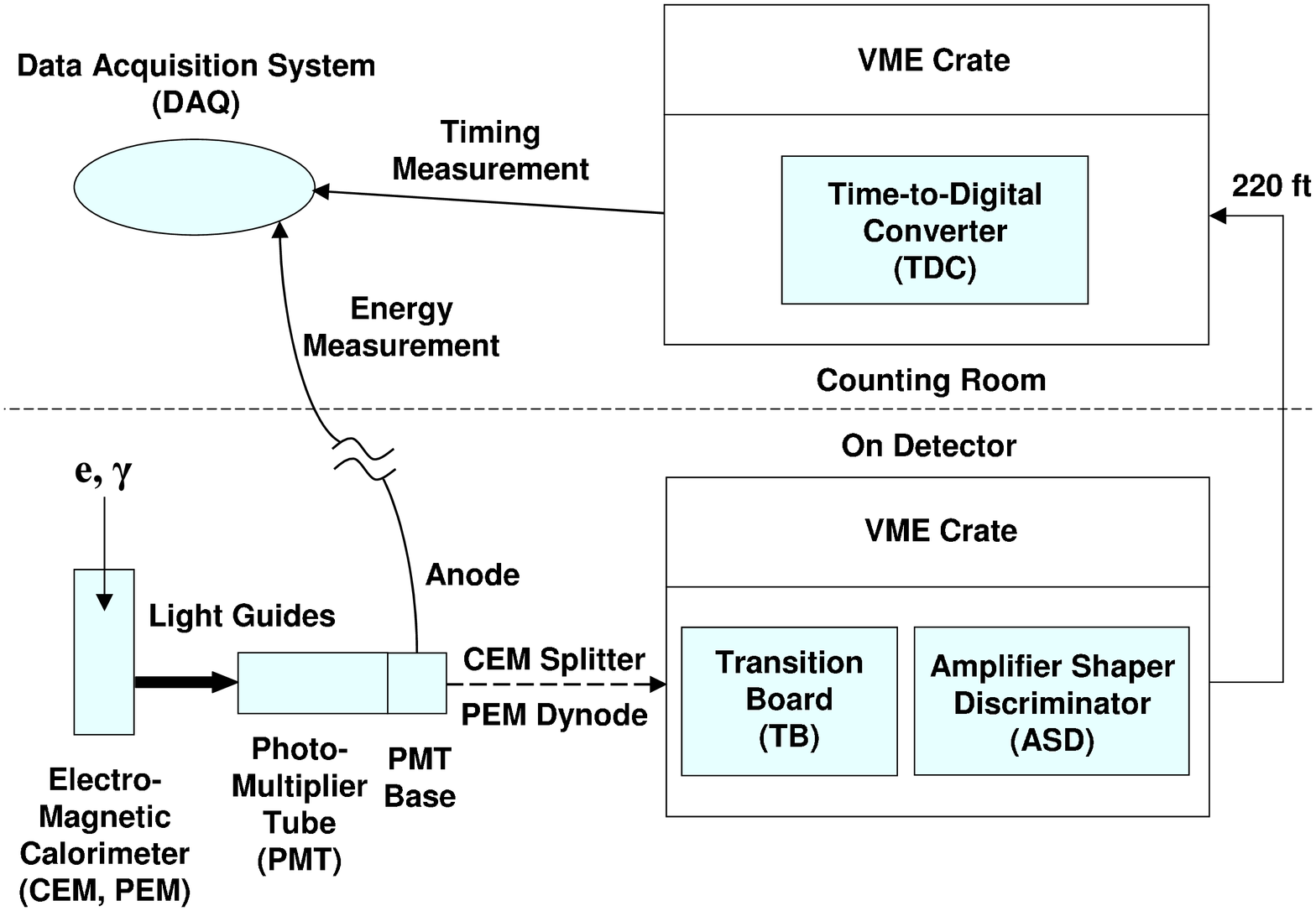}}
\caption{A schematic diagram of the EMTiming system hardware on the CDF detector.  Note that the CEM PMTs use ``splitter" outputs (described in Section~\protect\ref{PMTs}) to send their signals to the discriminators through transition boards, while the PEM uses dynode outputs. }
\label{fig:gemtime}
\end{figure}
\end{center}
}

\newcommand{\FigSplitterDiag}{
\begin{center}
\begin{figure}[bp]

\centering{\includegraphics[viewport=00 0 500 320,clip,width=1.0\linewidth]{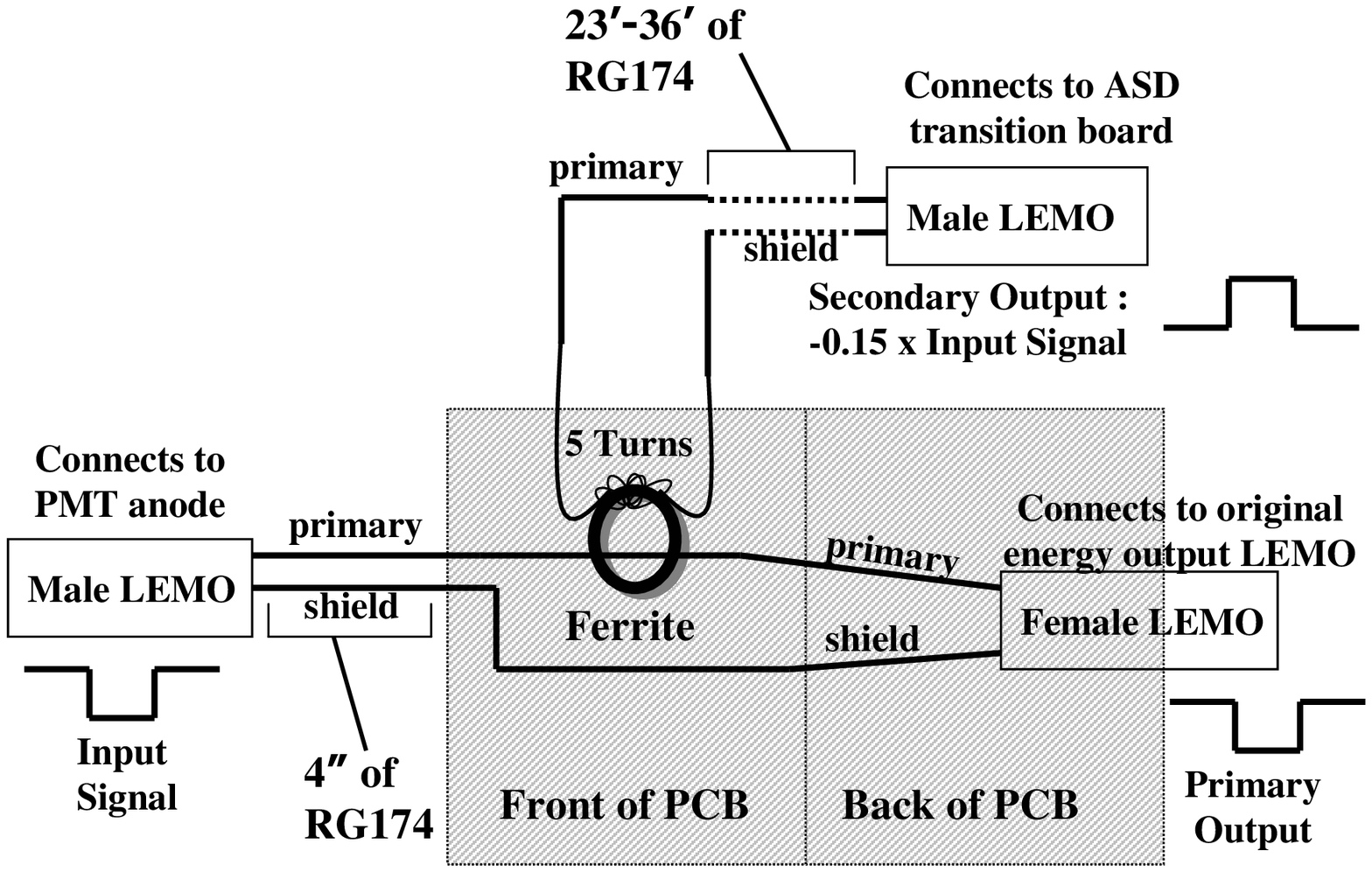}}

\caption{An electronics schematic for the ``splitter" printed circuit board (PCB) used in the CEM timing measurement. Note that RG174 refers to the coaxial cable used in the system, and that the primary output is used for the energy measurement while the secondary output is used for the timing measurement and is the input to the transition boards.  }
\label{fig:splitter_diagram}
\end{figure}
\end{center}
}

\newcommand{\FigASD}{

\begin{figure}[htbp]
\centering{\includegraphics[viewport=10 40 710 350,clip,width=1.0\linewidth]{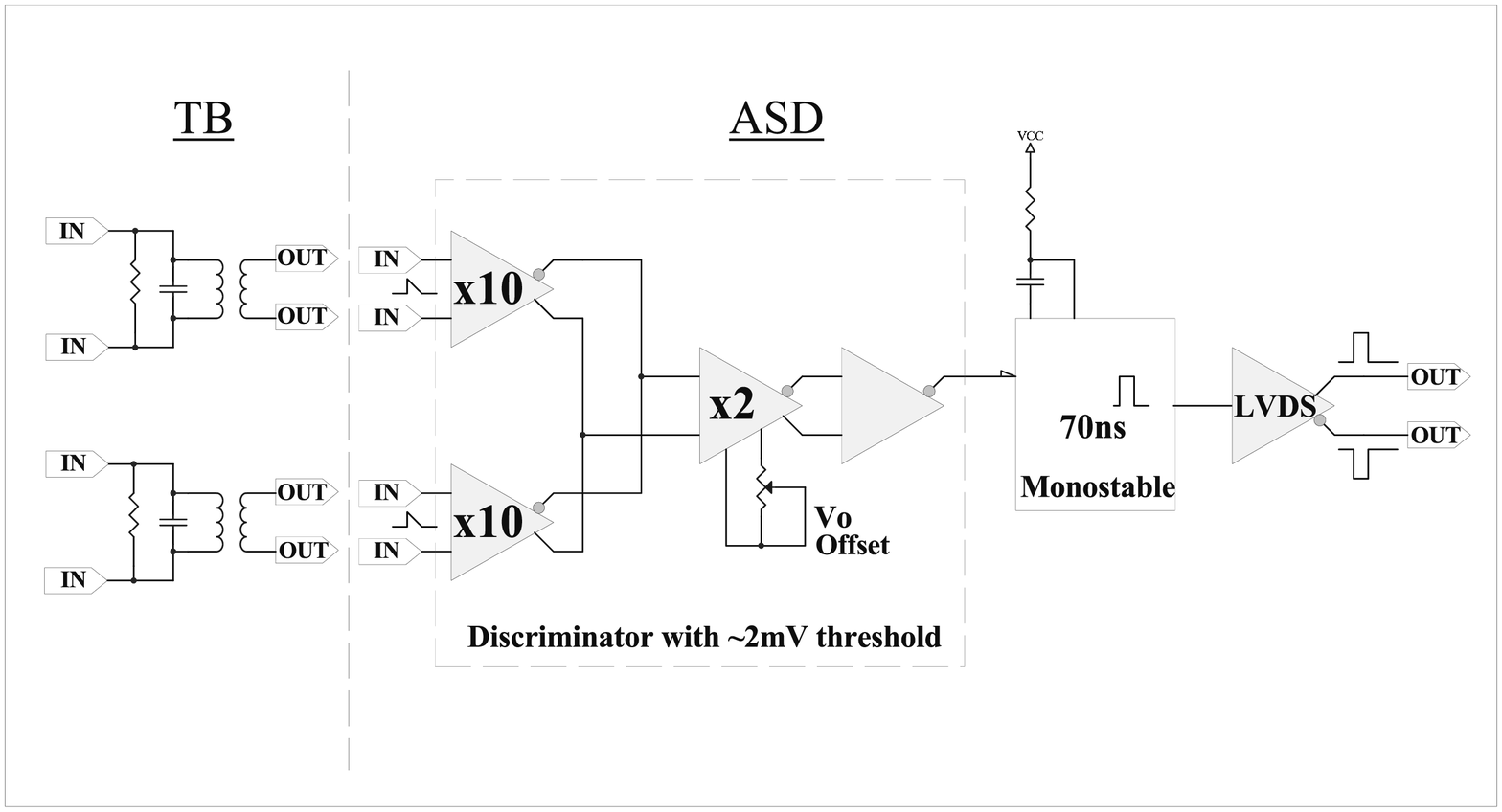}}
\caption{A schematic for the signal processing that occurs on the Transition Board (TB) and Amplifier-Shaper-Discriminator board (ASD). The $RC$ circuit ($R=150~\Omega$, $C=$12~pF) in parallel with a transformer (1:1) on the TB is designed to reduce noise and reflection problems at the input ($Z=50~\Omega$). On the ASD the amplifiers and comparator effectively sum the two PMT signals and discriminate on the leading edge with a 2~mV threshold. }
\label{fig:asd_scheme}
\end{figure}
}

\newcommand{\FigSysEff}{

\begin{figure}[htbp] 
\centering{
\psfig{file=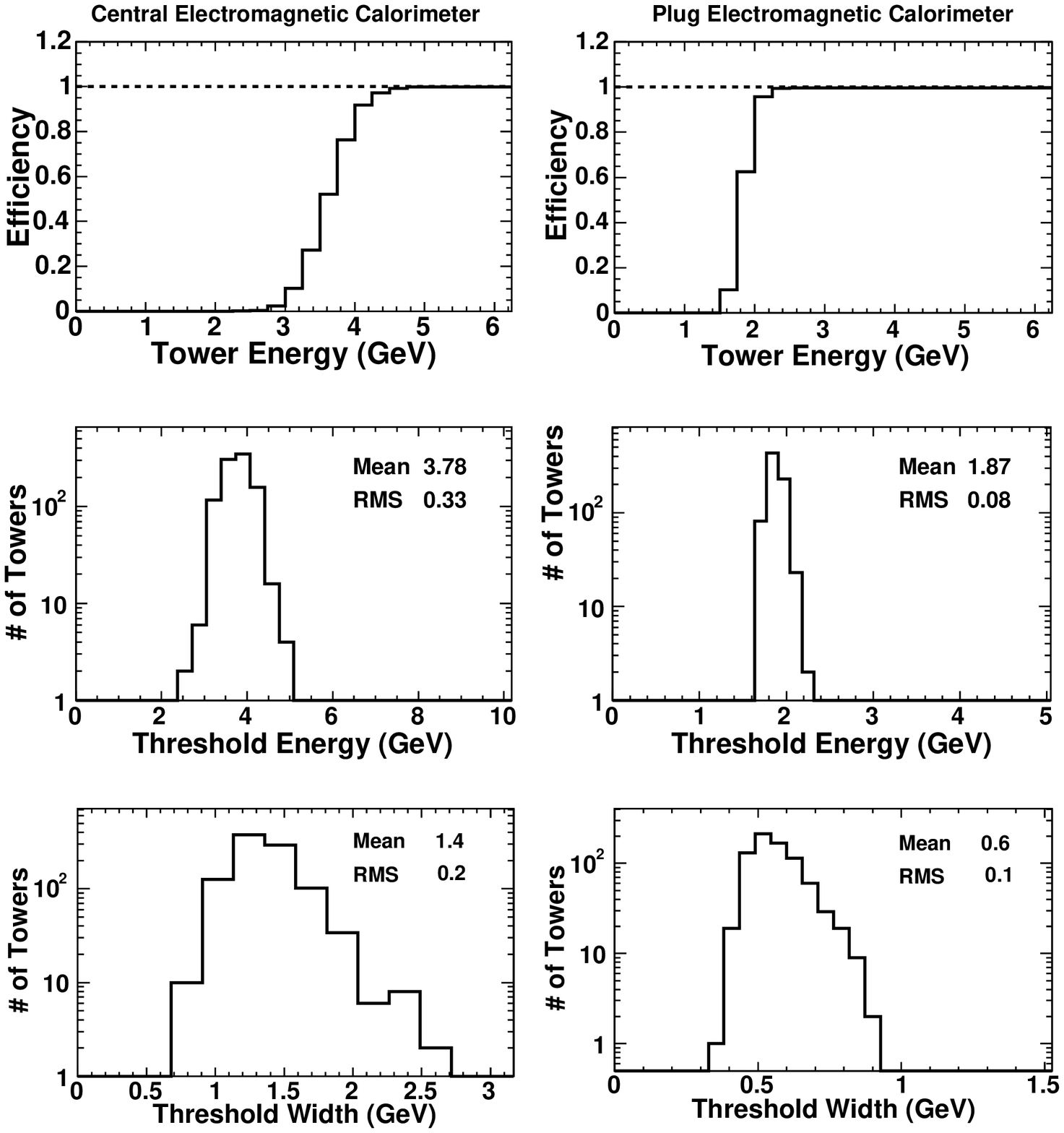,width=1.0\linewidth}}
\caption{The EMTiming system response as a function of the energy deposited in the EM calorimeter for a sample of hadrons from jets measured in the detector. The top plots show the efficiency (the fraction of events with a time recorded in the TDC to all events) as a function of the energy for the CEM and PEM, and includes all channels. The bottom two rows show histograms of the energy threshold and threshold width of the individual channels and indicate the uniformity of the system.}
\label{fig:sysefficiency}
\end{figure}
}

\newcommand{\FigSysResZ}{
\begin{figure}[tbp]
\vspace{-0.6cm}
  \centering{
  \psfig{file=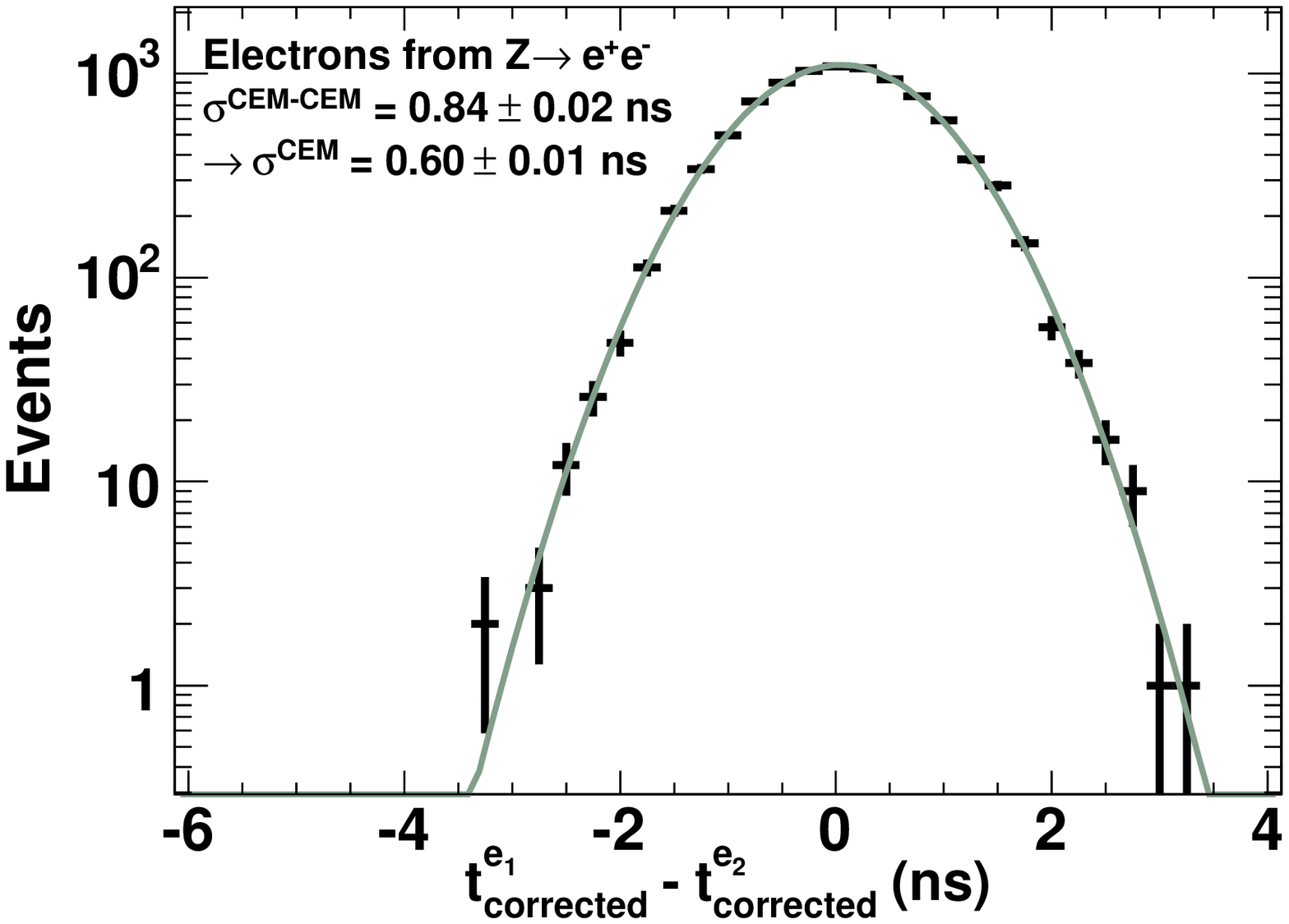,width=1.0\linewidth}
  \psfig{file=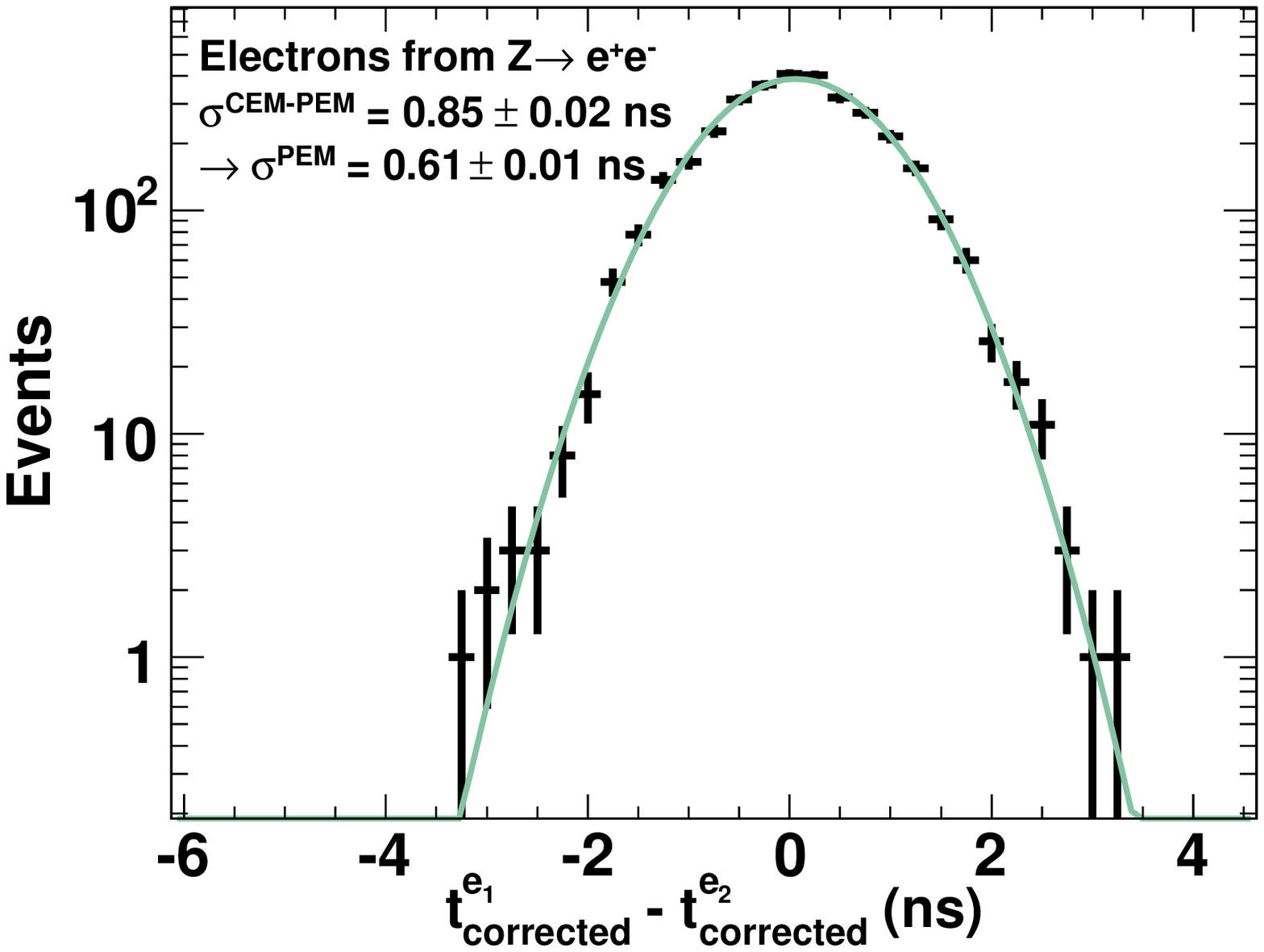,width=1.0\linewidth}
  \psfig{file=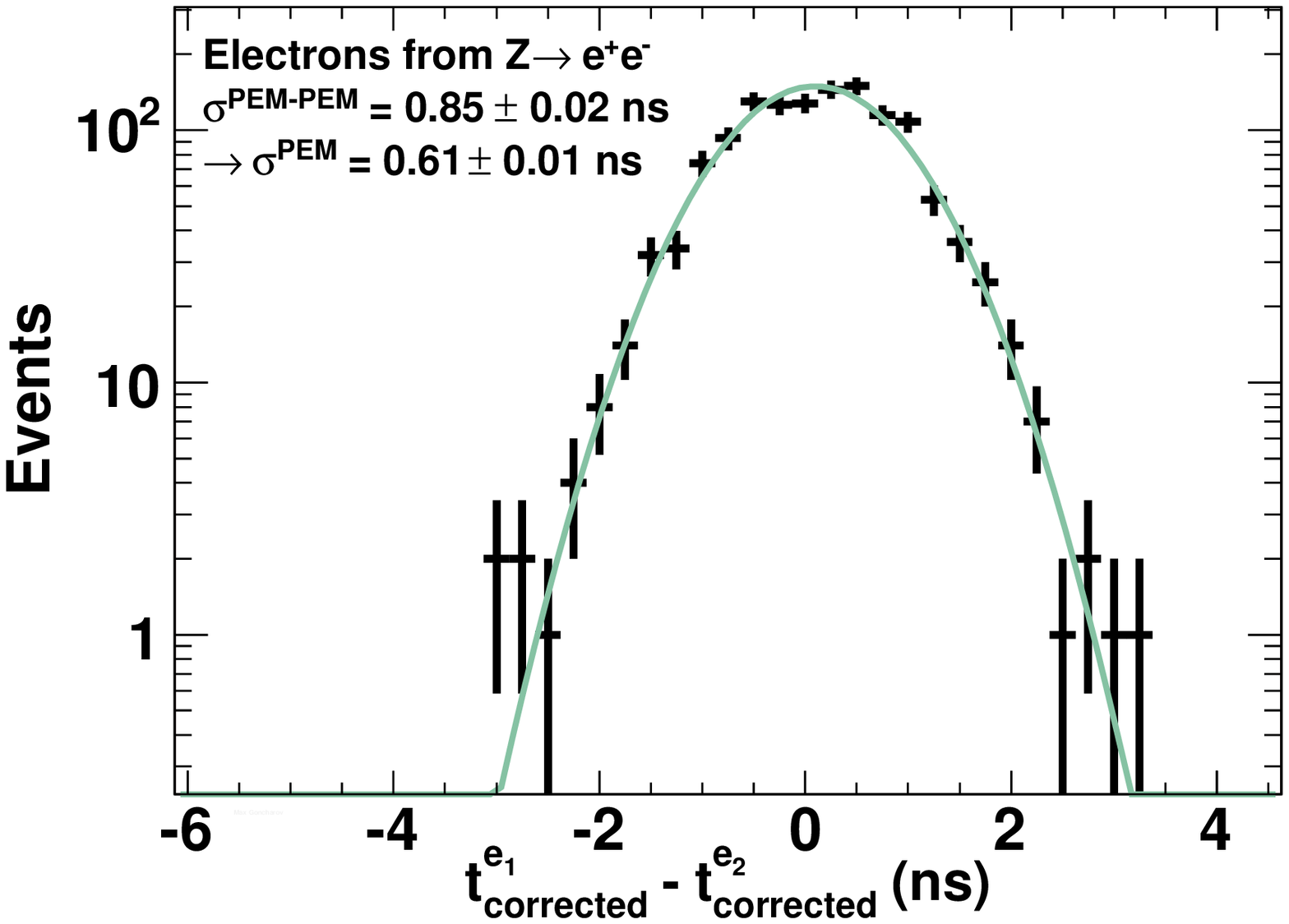,width=1.0\linewidth}
\vspace{-0.6cm}
}
\caption{The difference between the measured times of two electrons from three different samples of $Z\rightarrow ee$ events (CEM-CEM, CEM-PEM and PEM-PEM from top to bottom respectively). The CEM and PEM resolutions can be determined by taking the RMS/$\sqrt{2}$ which gives a single electron resolutions of 600$\pm$10~ps and 610$\pm$10~ps respectively. }
\label{fig:resolutionZ}
\end{figure}    

}

\newcommand{\FigSysResW}{
\begin{figure}[tbp]
\vspace{-0.6cm}
  \centering{
  \psfig{file=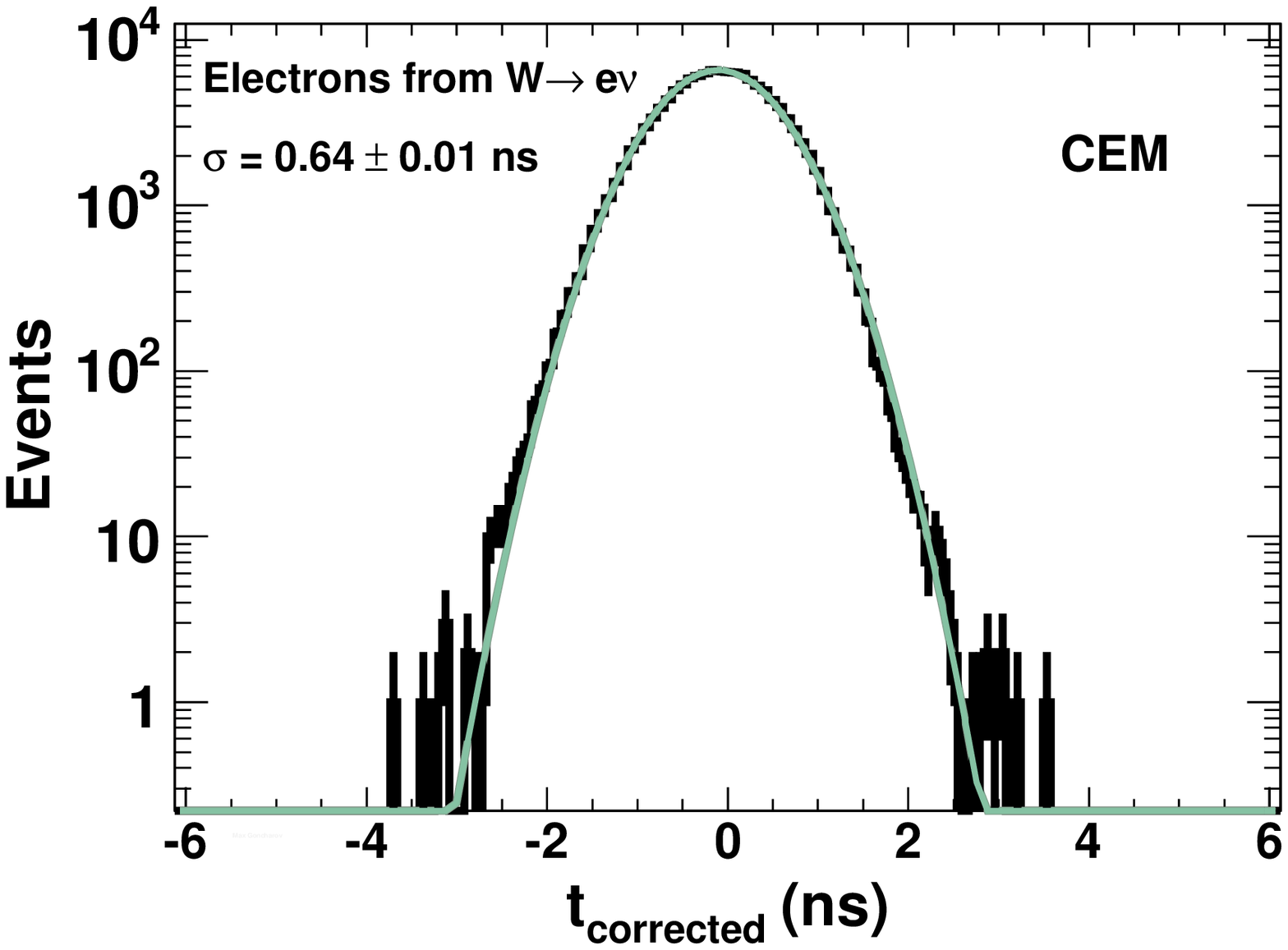,width=1.0\linewidth}
  \psfig{file=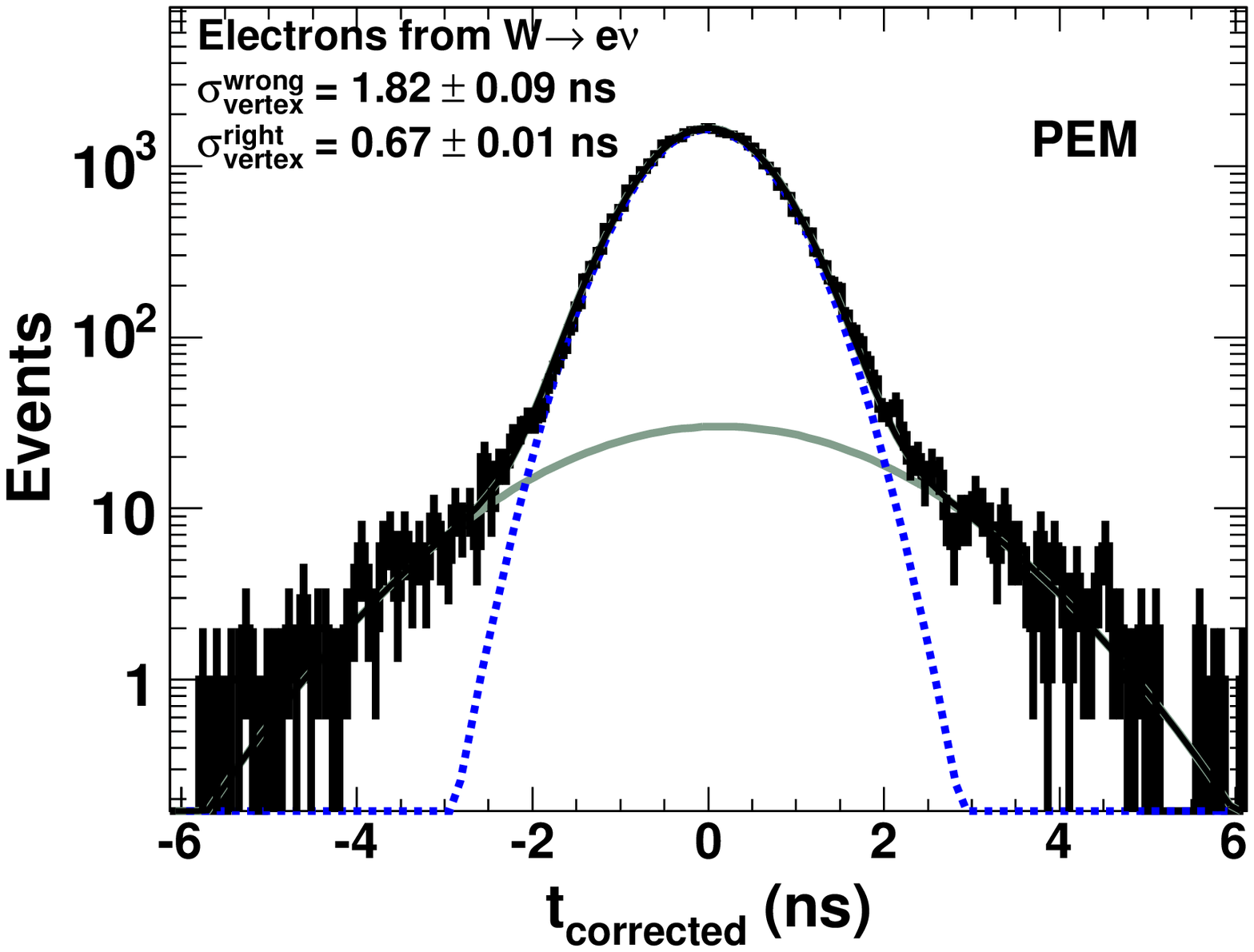,width=1.0\linewidth}
\vspace{-0.6cm}
}
\caption{The timing resolution for single electrons as measured with a sample of $W \rightarrow e\nu$ events. Note that the RMS of these distributions is slightly larger than in fig.~\protect\ref{fig:resolutionZ} as it includes the resolution of the collision time measurement that is different in the CEM and PEM. The second Gaussian in the PEM distribution corresponds to cases where there is a second collision in the event that was incorrectly selected as the event collision time as discussed in the text.}
\label{fig:resolutionW}
\end{figure}    

}

\newcommand{\FigSysResEner}{
\begin{figure}[htbp]
  \centering{
  \psfig{file=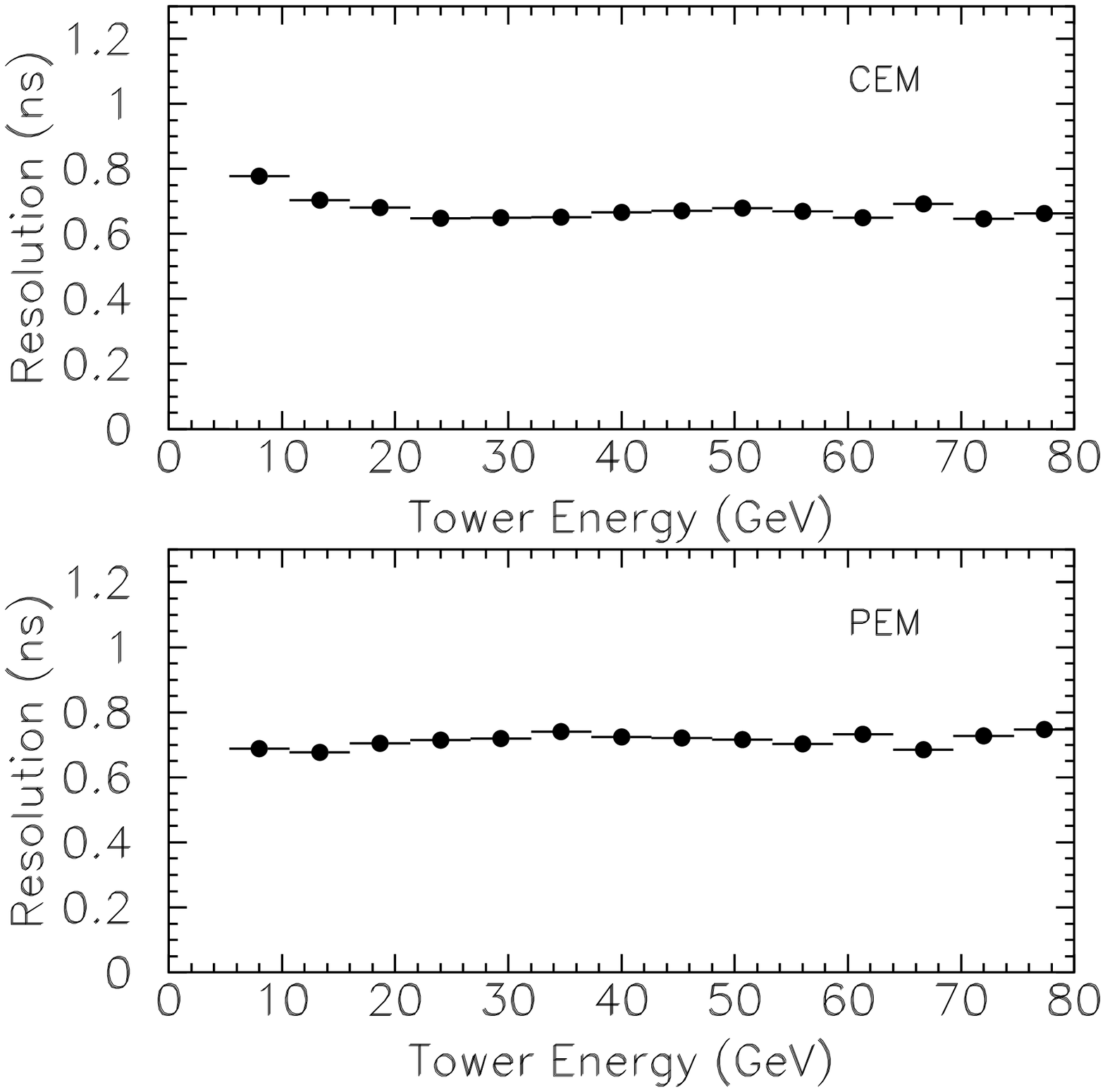,width=1.0\linewidth}
\vspace{-0.6cm}
}
\caption{The timing resolution as a function of energy. Note that the resolution, as in fig.~\protect\ref{fig:resolutionW}, contains a contribution from the uncertainty in the collision time that is different for the CEM and PEM. In each case the asymptotic value is the same as the electron resolution in fig.~\protect\ref{fig:resolutionW}.  }
\label{fig:resolutionEner}
\end{figure}    

}

\newcommand{\TabChannels}{

\begin{table*}
\begin{center}
\caption{Overview of the EMTiming system hardware and performance. Note that the calorimeter is physically segmented into 1,246 ``towers", and that each timing channel consists 
of two input PMTs for each ASD/TDC output line. As described in Section~\protect\ref{syseff} the efficiency for a time to be recorded is energy dependent and is well described by an efficiency plateau, threshold and width. All channels have a plateau which is $\approx$100\% efficient, and the uncertainties listed for the threshold and width values represent the 
channel-to-channel variation in the system. 
As described in Section~\protect\ref{sysres} the timing resolution asymptotically improves as a function of energy, and the numbers quoted here are for energies above 10~GeV and 6~GeV for the CEM and PEM respectively. For more details on the CEM and PEM calorimeters see 
Refs.~\protect\cite{CEM,PEM}. 
}
\vspace{0.1in}
\begin{tabular}{lll}
\hline
				& \multicolumn{1}{c}{CEM} 			& \multicolumn{1}{c}{PEM}              	\\ 
Coverage			& $|\eta|<1.1$ 		& $1.1<|\eta|<2.1$	\\ 
PMT				& Hamamatsu R580 	& Hamamatsu R4125   	\\ 
Physical tower segmentation 	& $\Delta\phi = 15^{\deg}$, $\Delta\eta \approx 0.1$ & 
				$\Delta\phi = 7.5^{\deg}$, $\Delta\eta \approx 0.1$ \\ 

Tower readout  			& 2 PMTs per tower 
				& 1 PMT per tower \\
PMT$\rightarrow$ ASD readout method
				& \begin{minipage}{2.1in} Analog Splitter, both PMTs from a single tower combined\end{minipage}	
				& \begin{minipage}{2.1in} Dynode, PMTs from two adjacent towers combined   \end{minipage}	           	\\ 
Total number of PMTs/ASD channels 
& 956/478 		& 768/384           	\\ 

Number of TB/ASD/TDC boards 
		& 32/32/8 		& 16/16/4	\\ 
Energy threshold (50\% efficiency point) 		& 3.8$\pm$0.3 GeV	& 1.9$\pm$0.1 GeV     	\\ 
Threshold width						& 1.4$\pm$0.2~GeV 	& 0.6$\pm$0.1 GeV      	\\ 

Timing resolution at asymptotic energies 	& 600$\pm$10 ps & 610$\pm$10 ps          \\ \hline
\end{tabular}
\label{tab:channels}
\end{center}
\end{table*}
}

\usepackage{epsfig}
\usepackage{calc,rotating,xspace}

\def\and{\\ \vspace{0.3cm}}

\begin{document}
\oddsidemargin 0.5in
\evensidemargin 1.5in
\begin{frontmatter}

\title{The Timing System for the CDF Electromagnetic Calorimeters}
\author{M.~Goncharov},
\author{T.~Kamon},
\author{V.~Khotilovich},
\author{V.~Krutelyov},
\author{S.W.~Lee},
\author{D.~Toback},
\author{P.~Wagner}
\address{Texas A\&M University}
\author{H.~Frisch},
\author{H.~Sanders}
\address{University of Chicago} 
\author{M.~Cordelli},
\author{F.~Happacher},
\author{S.~Miscetti}  
\address{INFN-Frascati}
\author{R.~Wagner}
\address{Argonne National Laboratory}

\date{\today}

\thispagestyle{empty}
\maketitle


\begin{abstract}

We report on the design and performance of the electromagnetic calorimeter timing readout system (EMTiming) for the Collider Detector at Fermilab (CDF). 
The system will be used in searches for rare events with high energy photons to verify that the photon is in time with the event collision, to reject cosmic ray and beam halo backgrounds, and to allow direct searches for new heavy long-lived particles that decay to photons.
The installation and commissioning of all 862 channels
was completed in Fall 2004 
as part of an upgrade to the Run~II version of the detector. 
Using in-situ data, including electrons from $W\rightarrow e\nu$ and $Z \rightarrow ee$ decays, we measure the energy threshold for a time to be recorded  to be 3.8$\pm$0.3~GeV in the central portion of the detector and 1.9$\pm$0.1~GeV in the plug portion. 
Similarly, we measure a timing resolution of 600$\pm$10~ps and 610$\pm$10~ps for  electrons above 10~GeV and 6~GeV respectively. There are very few system pathologies such as recording a time when no energy is deposited, or recording a second, fake time for a single energy deposit.







\end{abstract}


\end{frontmatter}
\clearpage
\clearpage

\newcounter{myctr}


\section{Introduction}
\label{sec:introduction}

\textwidth 7.5in 
\oddsidemargin -0.5in

\mysect{Overview of CDF and the CDF Calorimeters}

Timing readout for the electromagnetic (EM) calorimeters~\cite{CEM,PEM} was recently installed as part of an upgrade to the Run~II version of the Collider Detector at Fermilab (CDF)~\cite{CDF}. This system, known as \mbox{EMTiming}~\cite{Webpage}, is similar to the existing hadronic calorimeter system~\cite{HADTDC} but has a resolution of less than a nanosecond
and covers the central (CEM, $|\eta|<1.1$) and 
plug (PEM, $ 1.1<|\eta|<2.1$) portions of the calorimeter, where $\eta = -{\rm ln~tan}(\theta/2)$, and $\theta$ is the angle
from the beamline.

\mysect{Quick physics motivation: 3 Reasons}

The design of the EMTiming system is optimized for searches for production of new particles that decay into high energy photons as would be the case in some models of
Supersymmetry or Large Extra Dimensions~\cite{NewPhysics}. Final state particles from all proton anti-proton collisions typically traverse the CDF detector and the EMTiming system records the time of arrival for those that deposit large amounts of energy in the EM calorimeter. To improve the search sensitivity and robustness, the system can verify that all 
photons (or other energy) are from the primary collision~\cite{eeggMet} and reject and estimate the rate of cosmic ray 
and beam-related backgrounds. In addition, the system also allows for a new class of searches for heavy, long-lived particles that decay to photons via their delayed time of arrival~\cite{Wagner}. 

In the following sections we describe the various components of the EMTiming 
system and the system performance from its first four months of operation. In Sections~\ref{Overview} - \ref{ASD} we give an overview of the system and describe the path of the calorimeter signals through the various components in the system, with particular emphasis on the differences between the CEM and PEM systems which were originally built with different readout. In Section~\ref{sysperf} we describe the preliminary system performance as measured 
using collision data in-situ.

\section{Overview of the System}
\label{Overview}

\mysect{System Functionality and Hardware Design Considerations}

Particles from the collision that deposit energy in the EM calorimeters create light within the scintillators~\cite{CEM,PEM} that can be used for a timing measurement. Photo-multiplier tubes (PMTs) collect this light and effectively
convert the particle energy into an analog signal. Prior to the installation of the EMTiming system this signal was only used as part of the energy measurement. The EMTiming system, shown schematically in 
fig.~\ref{fig:gemtime}, routes a copy of the PMT signal to a passive Transition Board (TB) and 
Amplifier-Shaper-Discriminator board (ASD) pair that, in turn, sends their signal
to a fixed-start Time-to-Digital Converter (TDC) board for a timing measurement. 


The system design takes into account the differences between the existing CEM and PEM 
readout while trying to make the overall system as uniform and modular as possible. While the PEM, assembled for the beginning of Run~II, was designed with a timing upgrade in mind, the CEM was not. In Section~\ref{PMTs} we describe the differences 
between the PMT outputs for the CEM and PEM, and the designs used to get a 
timing signal out of the PMTs and onto the ASD boards. In Section~\ref{ASD} 
we describe the boards that collect the signals from the PMTs, amplify, shape, combine 
and use them as input to the discriminator, and output the results to a TDC. 
Table~\ref{tab:channels} provides a summary of the system 
specifications.


\FigOverview

\section{The CEM and PEM PMT Output Signals}
\label{PMTs}

\TabChannels

\mysect{Overview of the CEM and PEM PMTs}

As described in table~\ref{tab:channels}, the CEM and PEM have both different designs and PMT readouts. 
In the CEM, the original PMT bases are custom designed\footnote{We note that these bases were built almost 20 years prior to the EMTiming installation, and are largely unmodified since then.} and only 
provide an anode output via a LEMO connector. In the PEM, the 
system was designed with the dynode signal already available using AMP 
connector units.

\mysect{CEM: Splitter solution}

\FigSplitterDiag

In order to continue to use the pre-existing CEM hardware, we designed an inductive signal ``splitter" board 
that is placed between the PMT base and the original 25~ft RG174 cable that carries the anode signal for an energy measurement. The 
splitter effectively routes a fractional portion of the PMT pulse energy for timing use, while not affecting the 
energy measurement. As shown in fig.~\ref{fig:splitter_diagram} the anode cable is connected, via LEMO, to a printed circuit board where the primary and shield wires are separated. The primary line is then passed through a small circular ferrite after which the wires are rejoined.  A secondary wire is wound around the ferrite so that a signal from the PMT anode inductively generates a voltage for timing use. This secondary signal is sent via RG174 with a cable length that 
varies between 23 and 36~feet to the TBs depending on the PMT location on the detector. The splitter solution avoids potential
ground-loop problems since there is no electrical contact between the timing and 
energy readout lines. All cables have LEMO connectors on each end to facilitate installation.

The secondary pulse used for the timing measurement has a voltage that is 15\% of the primary signal. Since the two lines are only inductively coupled, and the energy 
measurement is done with a charge integration device~\cite{ADMEM}, in principle this solution should not affect the energy measurement since no charge can be lost. To test this, test stand comparisons of the integrated charge for a PMT pulse, with and without a splitter, were performed at various points over the full energy integration range. There was no observed (systematic or otherwise) effect on the linearity or resolution for all energies, with a measurement uncertainty of approximately 10\% of the 1$\sigma$ variation in the charge integration measurement itself (for a given energy).



The PEM was designed with a potential timing upgrade in mind. The 
dynode outputs from each PMT are collected in groups of 16 and made available 
for connection via off-the-shelf AMP connectors. The individual dynode outputs are then sent via 25~ft RG174 cables to the TBs and connected via LEMOs.


\section{Signal Discrimination and Transfer to the TDC}
\label{ASD}

\mysect{Signal path...TB}

Beginning with the transition boards the remaining system components, including the ASD boards, the long cables and the TDCs, are identical for the CEM and PEM. The number of boards for each sub-system is summarized in table~\ref{tab:channels}. 

Each TB is capable of taking in the cables from 48~PMTs. As shown in fig.~\ref{fig:asd_scheme}, processing begins with an $RC$ circuit and a transformer to both help reduce reflections back to the PMT that might otherwise cause a second, erroneous signal to be sent, and to reduce ground-loop or DC offset problems that might induce a signal to be sent when no energy is deposited. Each line is passed, via crate backplane, to the ASD. 

\mysect{ASD}
On the ASD the signals from two PMTs are combined and compared to a threshold with the digital result sent to the TDC. As shown on the right-hand side of fig.~\ref{fig:asd_scheme}, each PMT signal is amplified, then combined in an analog sum with a threshold offset, and sent to a comparator. This effectively operates as a 
fixed-height discriminator with a 2~mV threshold. The resulting signal triggers a monostable with an output width of 70$\pm$10~ns that in turn controls a National Semiconductor DS90C031 LVDS 
driver.




 
 




The output for each channel on an ASD board is placed on a single 220~ft multiwire twisted pair cable (3M 
3644B/50) that goes from the collision hall calorimeter crates upstairs to 
the TDCs. At this length these cables have a rise time of $\sim$50~ns. 
However, test bench studies show that as long as the input LVDS signal width is 
$>$50~ns, then we expect negligible data-transfer loss and timing jitter of 
$<$30~ps. The receiver chip set is located on one of the standard CDF 96~channel TDCs~\cite{TDC} which are a part of the data acquisition system (DAQ) and readout for each event.


\FigASD

%

\section{System Performance}
\label{sysperf}

\mysect{Overview and datasets}
The EMTiming system operates at CDF when the Tevatron is in Collider mode. During data taking protons and anti-protons collide every 396~ns, on average, at the center of the detector. The variation in the time and position of the collision is $\sigma_{time} \approx$1.3~ns and $\sigma_{position} \approx$25~cm, respectively. The typical time of flight from the collision point to the calorimeter is 5~ns and has an RMS variation of about 0.6~ns. If there is enough energy deposited, the TDC will record a time of arrival and this time is straightforwardly corrected for each of the above, and other, effects.

The system operation can be described in terms of its efficiency to record a time, the timing resolution, and the rate 
of pathologies. We study the performance in-situ using collisions that produce hadrons from jet events, and electrons from $Z \rightarrow ee$  and $W\rightarrow e\nu$ events. The results presented here represent the data 
taken during the first four months of Tevatron running in 2005.

\subsection{System Efficiency}
\label{syseff}
\mysect{System Efficiency}

The efficiency, the ratio of events with a time recorded in the TDC to all events, is a strong function of the PMT output signal size and is thus energy dependent. The efficiency as a function of the energy deposited in the tower is shown in the top part of fig.~\ref{fig:sysefficiency} for the CEM and PEM separately, and includes all towers together. The distribution is well 
described by an error/smeared step function, $Erf(E_{th},\sigma ,\epsilon )$, where $E_{th}$ is the 
threshold, $\sigma$  is the transition width at threshold, and $\epsilon$  is the 
plateau efficiency. We investigated each fully instrumented tower separately and find that the efficiency plateaus at 100\% and, as shown in the bottom part of fig.~\ref{fig:sysefficiency}, that the threshold and width values, as determined from a fit, are quite uniform. 
Table~\ref{tab:channels} summarizes the results. 

\FigSysEff

\subsection{Timing Corrections and System Resolution}
\label{sysres}
\mysect{Timing Corrections and System Resolution}

The time of arrival recorded by the TDC is a ``raw" time and has a 8~ns variation due to different effects. We take into account the dominant effects to produce a ``corrected time" distribution that is centered at 0~ns for relativistic particles promptly produced in the collision~\cite{Webpage}. The corrected time is given by:

\begin{eqnarray}
\rm{
t_{corrected}} & = & \rm{t_{Raw} + C_{Start} + C_{Energy} + C_{Energy~Asymmetry}} \nonumber \\
  & & \rm{- C_{Time~of~Flight} - C_{Collision~Time} , }
\end{eqnarray}

\noindent where the corrections are described below and determined on a tower-by-tower basis, unless otherwise noted, using in-situ data. 

\begin{itemize}

\item Since the system uses fixed-start TDCs there is a constant offset between the average time of arrival of the 
energy deposited in the calorimeter and the TDC start. This constant, ${\rm C_{Start}}$, is dominated by the overall cable lengths and thus can vary greatly from tower-to-tower. 

\item Since the ASDs use a fixed-height discrimination method the PMT pulse shape produces a ``slewing" that makes the recorded time energy dependent. We use a set of empirically derived constants given by:

\begin{equation}
\hspace{0.8in}
\centering{{\rm C_{Energy}} = \frac{A_1}{ln(x)} + \frac{A_2}{x^2}}
\end{equation}

\noindent where $A_1$ and $A_2$ are constants and $x$ is the sum of the energies from the two PMTs, as 
measured in the calorimeter. 

\item PMT energy response differences as well as the location within a tower where the particle hits can also affect the measured time of arrival. We correct for both effects by taking into account the energy asymmetry between the towers. We use the empirically derived functional form: 

\begin{equation}
\hspace{0.3in}
{\rm C_{Energy~Asymmetry}} = B_0 + B_1\cdot x + B_2\cdot x^2, 
\end{equation}

\noindent where $B_0, B_1$ and $B_2$ are constants and $x$ is the energy asymmetry of the two PMTs. 
After the above corrections the timing distribution is roughly a Gaussian with an RMS of ~1.6 ns. 

\item The arrival time is corrected for the expected time-of-flight, ${\rm C_{Time~of~Flight}}$, using
the measured collision position from the tracking chamber~\cite{COT} and the known calorimeter tower position. 


\item Finally, the measured collision time, ${\rm C_{Collision~Time}}$, as measured by the tracking chamber, is subtracted off on 
an event-by-event basis.

\end{itemize}

The system resolution is measured two ways. Using $Z \rightarrow ee$ events we subtract the corrected times of the two electrons, thus canceling the collision time and reducing any other global event mismeasurements effects. The result is shown in fig.~\ref{fig:resolutionZ} for combinations of electrons in the CEM-CEM, CEM-PEM and PEM-PEM respectively. The RMS per electron is 600$\pm$10~ps and 610$\pm$10~ps for the CEM and PEM respectively, and is consistent between the measurements. This includes the irreducible TDC resolution of 288~ps (the TDC has a 1~ns binning output, which corresponds to a 1~ns/$\sqrt{12}$ = 288~ps) as well as a negligible contribution to the uncertainty on C$_{\rm Time~of~Flight}$ from the vertex position measurement. 

\FigSysResZ

Fig.~\ref{fig:resolutionW} shows the timing resolution as estimated from a higher statistics sample of $W\rightarrow e\nu$ events where the collision time is measured, and corrected for, directly. While the results are consistent with the distributions in fig.~\ref{fig:resolutionZ} and show that the timing measurement is well described by a Gaussian even out to roughly 5$\sigma$ for the CEM, a few notes are in order to explain some of the differences. 
The RMS of the distributions are slightly larger as they include the resolution of the collision time measurement that is slightly larger for the PEM than for the CEM. In both distributions we have required a tight match between the electron track and the measured collision time and position. 
However, while the time and position for the electron track can be measured in the CEM region, only a position measurement is available for tracks in the region covered by the PEM. Thus, in the bottom part of fig.~\ref{fig:resolutionW} there is a second Gaussian in the distribution that corresponds to the case where there are two collisions in the event that occur at the same position, but well separated in time; the tracking for the PEM is unable to distinguish which is the correct one. This second Gaussian corresponds, in this case, to 4\% of the events, but should be sample dependent, and has an RMS that reflects the association of the time of arrival with a random collision time. 

Fig.~\ref{fig:resolutionEner} shows the single tower energy resolution (convoluted with the vertexing time resolution as in  fig.~\ref{fig:resolutionW}) as measured from a mixed sample of jets and electrons. While the resolution is slightly worse for low energies, the asymptotic resolution stabilizes at energies just above 10~GeV and 6~GeV in the CEM and PEM respectively to values that are very close to the electron resolution from $W$'s.



\FigSysResW
\FigSysResEner


\subsection{Pathologies}
\mysect{System Fake Firings, Double Firings and Other problems}
Finally, we note the rate of system pathologies. We measure the rate at which the system records a time when there is no energy recorded (the fake rate)
by the DAQ to be typically less than one occurrence in 10$^8$ collisions, per tower. We note however that this rate can rise to one in 10$^7$ if we allow cases where there is some evidence that the energy is not properly readout by the DAQ. The rate at which multiple times are recorded for a single energy deposit, presumably from reflections, is
roughly the same as the fake rate, but can be as high as one in 10$^5$ collisions for a few ($<$10) towers. However, this is easily corrected for in the final data analysis.



\section{Conclusions}
\label{conclusions}

\mysect{Conclusions}
The EMTiming system provides timing readout for the CDF electromagnetic calorimeters with good uniformity and resolution. It has its 50\% efficiency points 
at 3.8$\pm$0.3~GeV and 1.9$\pm$0.1~GeV in the CEM and PEM respectively, and is 100\% efficient well above threshold. After a full set of corrections we find 
600$\pm$10~ps and 610$\pm$10~ps timing resolutions, respectively, with only small deviations from a Gaussian distribution. 
There are very few pathologies observed in the data such as recording a time when no energy is deposited or recording multiple times for a single energy deposit. The system is well understood and ready for searches for new particles that decay to photons.


\section{Acknowledgments}
The authors would like to acknowledge a number of people who contributed significantly to the EMTiming system. They include
Dervin Allen,  
Chris Battle,
Myron Campbell,
Matt {\mbox Cervantes,}
Steve Chappa, 
Jamie Grado, 
Bob DeMaat, 
Camille Ginsburg, 
Eric James,
Min-Jeong Kim,
Steve Kuhlmann, 
Jonathan Lewis,
Mike Lindgren, 
Pat Lukens,
Lew Morris, 
Peter Onyisi,
Steve Payne,
Fotis Ptohos, 
Rob Roser, 
Willis Sakumoto,
Paul Simeon,
Rick Tesarek,
Erwin Thomas,
Wayne {\mbox Waldrop,}
and
Peter Wilson. 
In addition, we thank the Fermilab staff and the technical staffs of the participating CDF institutions for their vital contributions. This work was supported by the U.S. Department of Energy and National Science Foundation; the Italian Istituto Nazionale di Fisica Nucleare; the Ministry of Education, Culture, Sports, Science and Technology of Japan; the Natural Sciences and Engineering Research Council of Canada; the National Science Council of the Republic of China; the Swiss National Science Foundation; the A.P. Sloan Foundation; the Bundesministerium f\"ur Bildung und Forschung, Germany; the Korean Science and Engineering Foundation and the Korean Research Foundation; the Particle Physics and Astronomy Research Council and the Royal Society, UK; the Russian Foundation for Basic Research; the Comisi\'on Interministerial de Ciencia y Tecnolog\'{\i}a, Spain; in part by the European Community's Human Potential Programme under contract HPRN-CT-2002-00292; the Academy of Finland; and the College of Science at Texas A\&M University.




\begin{thebibliography}{99}


\bibitem{CEM}

L.~Balka \etalnoem, \NIM{267}{1988}{272}. 


\bibitem{PEM}

M.~Albrow \etalnoem, \NIM{480}{2002}{524}.

\bibitem{CDF} 
R. Blair et al., CDF Collaboration, CDF II Technical Design Report, FERMILAB-Pub-96/390-E (1996). 

\bibitem{Webpage} \begin{sloppypar}For a more complete description of the EMTiming system as well as the most up to date reconstruction software see hepr8.physics.tamu.edu/hep/emtiming. \end{sloppypar}

\bibitem{HADTDC} A short description of the Run I hadron calorimeter timing (HADTDC) system can be found in S.~Bertolucci \etalnoem, \NIM{267}{1988}{301}.




\bibitem{NewPhysics} There are many models of new physics that predict photons in the final state. These include
%
%
S.~Ambrosanio \etal, \Journal{\PRL}{76}{3498}{1996};
%
S. Dimopoulos, M. Dine, S.~Raby, and S.~Thomas, \Journal{\PRL}{76}{3494}{1996};
J.~L.~Lopez and D.~V.~Nanopoulos, Mod.\ Phys.\ Lett.\ A {\bf 10} (1996) 2473; 
S. Ambrosanio, G.~L.~Kane, G.~D.~Kribs, S.~Martin, and S.~Mrenna, \Journal{\PRD}{54}{5395}{1996};
B.~C.~Allenach, S.~Lola, and K.~Sridhar, \Journal{\PRL}{89}{011801}{2002};
J. L. Rosner, \Journal{\PRD}{55}{3143}{1997};
%
%
U.~Baur, M. Spira, and P.~M.~Zerwas, \Journal{\PRD}{42}{815}{1990}; E.~Boos, A.~Vologdin, D.~Toback and J.~Gaspard, \Journal{\PRD}{66}{013011}{2002}; 
%
%
%
A.~D.~De~Rujula, M. B. Gavela, P. Hernandez, and E. Masso, Nucl.\ Phys.\ B {\bf 384} (1992) 3;
%
M. Baillargeon \etal,     hep-ph/9603220 (1996);
%
N.~Arkani-Hamed, S.~Dimopoulos, and G.~Dvali,  Phys. Lett. {\bf B429} (1998) 263; 
A.~Brignole, F.~Feruglio, M.~L.~Mangano, and F.~Zwirner, Nucl.~Phys.~B {\bf 526} (1998) 136; Erratum-ibid. B {\bf 582} (2000) 759;
%
H. Haber, G. Kane, and T.~Sterling, Nucl. Phys. B {\bf 161} (1979)  493;
A.~Stange, W.~Marciano, and S.~Willenbrock, \Journal{\PRD}{49}{1354}{1994};
M.G.~Diaz and T. Weiler, hep-ph/9401259 (1994); and
A.G. Akeroyd, Phys. Lett. {\bf B368} (1996) 89. 

\bibitem{eeggMet} Timing would have been very helpful for the interesting event described in F. Abe \etal\ (CDF Collaboration), \Journal{\PRL}{81}{1791}{1998} and \Journal{\PRD}{59}{092002}{1999}; D.~Toback, Ph.D. dissertation, Univ. of Chicago (1997).



\bibitem{Wagner} D.~Toback and P.~Wagner, \Journal{\PRD}{70}{114032}{2004}.




\bibitem{ADMEM} R.~Erbacher, Proc. of the 31$^{st}$ International 
Conference on High Energy Physics (ICHEP 2002), 
FERMILAB-CONF-02/251-E; 
T.~Shaw, C.~Nelson, and T.~Wesson, 
IEEE Trans. Nucl. Sci. 47 (2000) 1834. 

\begin{sloppypar}

\bibitem{TDC} {\mbox R.~Moore}, Proc. of the 2004 IEEE Nuclear Science Symposium and Medical Imaging Conference, 
\mbox{FERMILAB-CONF-04-262-E}. 

\end{sloppypar}

\bibitem{COT} T.~Affolder \etalnoem, \NIM{526}{2004}{249}. 



\end{thebibliography}
\end{document}